\documentclass[11pt,twoside]{jgsp}
\newcommand{\re}{\mathrm{e}}
\newcommand{\ri}{\mathrm{i}}
\newcommand{\rd}{\mathrm{d}}
\newcommand{\Tr}{\mathrm{Tr}}
\newcommand{\Fun}{\mathrm{C}}
\newcommand{\K}{\mathrm{K}}

\def\beq{\begin{equation}}
\def\eeq{\end{equation}}
\def\bea{\begin{eqnarray}}
\def\eea{\end{eqnarray}}

\newcommand{\R}{{\mathbb{R}}}
\newcommand{\N}{{\mathbb{N}}}
\newcommand{\C}{{\mathbb{C}}}
\newcommand{\Z}{{\mathbb{Z}}}
\newcommand{\Q}{{\mathbb{Q}}}
\newcommand{\M}{{\mathbb{M}}}
\newcommand{\T}{{\mathbb{T}}}
\newcommand{\Sphere}{{\mathbb{S}}}

\newcommand{\herm}{{\mathfrak u}}

\newcommand{\bff}{{\mbf f}}
\newcommand{\bfu}{{\mbf U}}
\newcommand{\bfphi}{{\mbf \Phi}}

\newcommand{\ds}{\displaystyle}

\newcommand{\proj}{{\sf P}}
\newcommand{\tach}{{\sf T}}

\newcommand{\hil}{{\cal H}}

\newcommand{\Lcal}{{\cal L}}
\newcommand{\alg}{{\cal A}}

\newcommand{\mbf}[1]{{\boldsymbol {#1} }}

\newcommand{\ncint}{\int\!\!\!\!\!\! - ~}

\newcommand{\one}{{1\!\!1}}
\def\Re{{\rm Re}\, }

\def\im{{\rm im}\, }

\def\ker{{\rm ker}\, }

\def\Tr{{\rm Tr}\, }

\def\aut#1#2{
{\small
\noindent
\parbox[t]{6.5cm}{#1}
 \hfill
\parbox[t]{6.5cm}{#2}
}
}

\begin{document}

\thispagestyle{plain}

\title{MATRIX MODELS, LARGE $\mbf N$ LIMITS AND NONCOMMUTATIVE SOLITONS}

\author{RICHARD J. SZABO}

\date{}

\maketitle

%\comm{Communicated by XXX}\
\noindent
\comm{{\small Based on plenary lecture given at
  the {\it XXIV Workshop on Geometric Methods in Physics}, Bialowieza,
  Poland, June 26 -- July 2 2005. To be published in {\sl Journal of
      Geometry and Symmetry in Physics}.}\\
{\small \tt HWM-05-28 , EMPG-05-18 , December 2005}}

\begin{abstract}
A survey of the interrelationships between matrix models and
field theories on the noncommutative torus is presented. The
discretization of noncommutative gauge theory by twisted reduced
models is described along with a rigorous definition of the large $N$
continuum limit. The regularization of arbitrary noncommutative field
theories by means of matrix quantum mechanics and its connection to
noncommutative solitons is also discussed. 
\end{abstract}

\label{first}

\section[]{Introduction\label{Intro}}

Two of the most novel aspects of noncommutative field theories
(see~\cite{KSrev1,DougNek1,Szrev1} for reviews) which
are not seen in ordinary quantum field theories are the properties
that (a) They can be regularized and analysed by means of matrix
models; and (b) In some instances they admit novel soliton solutions
with no commutative counterparts. Property (a) stems from the fact that
noncommutative fields are most conveniently understood and analysed as
operators acting on separable Hilbert spaces. Property (b) instead is
due to the fact that noncommutative field theories behave in many
respects like string theory, rather than conventional quantum field
theory, and noncommutative solitons correspond to {\it D-branes} in
open string field theory of tachyon dynamics (see~\cite{Solbrane1} for
a review). In
this survey we will describe some aspects of the interrelationship
between matrix models, noncommutative solitons and field theory on the
noncommutative torus. The first half of the article deals with the
finite-dimensional regularization of Yang-Mills theory on a
two-dimensional noncommutative torus~\cite{AMNSz1} and the precise
definition of the large $N$ limit of this matrix
model~\cite{LLSz1}. The second half demonstrates
that the two novel properties (a) and (b) above are in fact
intimately related through a regularization of noncommutative field
theory by means of matrix quantum mechanics which has a much simpler
and tractable large $N$ limit than its zero-dimensional
counterpart~\cite{lls2}.

\section{Matrix Models and Gauge Theory on the Noncommutative
  Torus\label{MMGTNT}}

We begin with an account of how the large $N$ limit of a
particular matrix model naturally leads to considerations of gauge
theory on a noncommutative torus. This is the classic
Connes-Douglas-Schwarz formalism~\cite{CDS1} which was the original
link between open string theory in background fields and
noncommutative geometry through compactifications of Matrix Theory.

\subsection{The IKKT Matrix Model\label{IKKT}}

Consider the statistical mechanics of $d\geq2$ complex $N\times N$
matrices $X_i\in\M_N(\C)$, $i=1,\dots,d$ which is defined by the
integral
\beq
{ Z}=\int_{\M_N(\C)\otimes\C^d}\rd X~\re^{-{ S}(X)}
\label{IKKTpartfn}\eeq
where $\rd X_i$ for each $i=1,\dots,d$ is the translationally
invariant Haar measure on the Lie algebra $\M_N(\C)$ and the action
\beq
{ S}(X)=-\frac1{g^2}\,\sum_{i,j=1}^d\,
\Tr\bigl[X_i\,,\,X_j\bigr]^2
\label{IKKTaction}\eeq
is a holomorphic function on the space $\C^d\times\M_N(\C)$. The
symbol $\Tr$ is the usual matrix trace and $g^2>0$ is a coupling
constant. Note that the Lie algebra $\M_N(\C)$ is also an associative
$*$-algebra.

The zero-dimensional matrix model (\ref{IKKTpartfn}) is simply the
dimensional reduction to a point of Yang-Mills gauge theory on $\R^d$
with structure group ${\rm GL}(N,\C)$, i.e. we restrict
the usual Yang-Mills action functional to {\it constant} gauge
fields. It possesses two fundamental symmetries. Regard
$X\in\M_N(\C)\otimes\C^d$ with components $X_1,\dots,X_d$ in a fixed
basis of the vector space $\C^d$. Then the function (\ref{IKKTaction})
is invariant under the action of the complex orthogonal group ${\rm
  SO}(d,\C)$ under which $X$ transforms as a vector. This symmetry is
just Lorentz invariance. The action is also invariant under the
transformations
\beq
X_i~\longmapsto~U\,X_i\,U^{-1} \ , \quad U\in{\rm GL}(N,\C) \ .
\label{Xigaugesym}\eeq
This symmetry is gauge invariance. We will restrict
the integral (\ref{IKKTpartfn}) over a real slice of the space
$\C^d\times\M_N(\C)$ by truncating to hermitean $N\times N$ matrices
$X_i\in{\mathfrak u}(N)$. Then the hermitean matrix integral
(\ref{IKKTpartfn}) defines the {\it IKKT matrix
  model}~\cite{IKKT1}. The gauge symmetry (\ref{Xigaugesym}) in this
instance restricts to unitary matrices $U\in{\rm U}(N)$.

This matrix model has many interesting applications which all
require some formal $N\to\infty$ limit to be taken. Firstly, identify
$\M_\infty(\C)$ as the $*$-algebra of operators of the form ${\cal
  D}+A$ with $\cal D$ a first order differential operator with
constant coefficients and $A$ an operator of multiplication on a
function which decreases rapidly at infinity in $\R^d$. The matrix model then
reduces to ordinary gauge theory on $\R^d$. Secondly, let ${\rm
  C}^\infty(\Sigma)$ be the algebra of smooth complex-valued functions
on a compact Riemann surface $\Sigma$ equiped with a symplectic
structure $\omega\in{\cal S}(\Sigma)$. Then one can construct a
sequence of ``quantization'' maps
$\sigma_N:\Fun^\infty(\Sigma)\to\M_N(\C)$ such that
$N\,[\sigma_N(f),\sigma_N(g)]\to\sigma_N(\{f,g\}_\omega)$ and
$N\,\Tr\,\sigma_N(f)\to\int_\Sigma\,\omega\,f$ in the limit
$N\to\infty$. Let $X:\Sigma\to\R^d$ be an embedding of the surface in
flat space. Then the extrema of the action $S(X)-\lambda\,{\rm
  vol}_\omega(\Sigma)$ regarded as a functional on
$\Fun^\infty(\Sigma)\times\C^d\times{\cal S}(\Sigma)$ coincide with
those of the Nambu-Goto action which computes the area of the embedded
surface $\Sigma$. This connects the IKKT matrix model to the
Green-Schwarz string~\cite{NBI1}. Finally, the action (\ref{IKKTaction})
itself
has a general set of classical vacuum states provided by
configurations $X_i$ with $[X_i,X_j]=\ri\,B_{ij}~\one$, where
$B_{ij}=-B_{ji}$ are real constants. For $B_{ij}\neq0$ these equations
only have a solution when $N\to\infty$ whereby $\M_\infty(\C)$ is
regarded as the $C^*$-algebra $\cal K$ of compact operators acting on
a separable Hilbert space.

These realizations all form part of the circumstantial evidence which
has led to the conjecture~\cite{IKKT1} that the large
$N$ limit of the matrix model (\ref{IKKTpartfn}) provides a
non-perturbative definition of Type~IIB string theory. The vacua
described above correspond to position coordinates of D-branes and
these considerations immediately lead to noncommutative spacetime
geometries~\cite{JapNC1,Wittenp}. The rigorous definition of the large
$N$ limit here along with the precise meaning of convergence will be
described later on.

\subsection{Toroidal Compactification\label{TorComp}}

We will now make the heuristic appearence of noncommutative geometry
described above more precise~\cite{CDS1}. Let us {\it compactify} the
space $\R^d$ along the $(12)$-plane to $\T^2\times\R^{d-2}$, where $\T^2$
is a square two-torus of sides $R_1$ and $R_2$. Since we interpret
the $N\times N$ hermitean matrices $X_i$ above as ``coordinates'' in
$\R^d$, we would like to define the toroidal compactication of the
IKKT matrix model. This means that we should define a restriction of
the action (\ref{IKKTaction}) to a subspace of ${\mathfrak u}(N)\otimes\R^d$
where an equivalence relation $X_i\sim X_i+2\pi\,R_i\,\one$, $i=1,2$
is satisfied, i.e. ${S}(X_i+2\pi\,R_i\,\one)={
  S}(X_i)$ for $i=1,2$. Using the gauge symmetry (\ref{Xigaugesym}),
we define this equivalence relation as unitary gauge equivalence and
hence consider the {\it quotient conditions}
\bea
X_i+2\pi\,R_i\,\one&=&U_i\,X_i\,U_i^{-1} \ , \quad i=1,2 \nonumber\\
X_j&=&U_i\,X_j\,U_i^{-1} \ , \quad \forall j\neq i \ , \ j=1,\dots,d \ ,
\ i=1,2
\label{quotientconds}\eea
where $U_1$ and $U_2$ are unitary, $U_i^{-1}=U_i^\dag$.

Taking the trace on both sides of the first equation in
(\ref{quotientconds}) shows that these
conditions cannot be satisfied by finite-dimensional matrices unless
$R_1=R_2=0$. Thus we take the formal large $N$ limit again and search
for solutions to these equations in terms of operators $X_j$, $U_i$ on
a separable Hilbert space $\cal H$. Consistency of the quotient conditions
requires that the object $U_1\,U_2\,U_1^{-1}\,U_2^{-1}$ commutes with
{\it all} $X_j$, $j=1,\dots,d$. A natural choice is then to set it
equal to a scalar operator on $\hil$,
$U_1\,U_2\,U_1^{-1}\,U_2^{-1}=\lambda\,\one$. Unitarity restricts
$\lambda=\re^{2\pi\,\ri\,\theta}$ for some $\theta\in\R$. We
thereby find that the unitary operators $U_1$ and $U_2$ obey the
relation
\beq
U_1\,U_2=\re^{2\pi\,\ri\,\theta}~U_2\,U_1 \ .
\label{NCtorusdef}\eeq
With the relation (\ref{NCtorusdef}), $U_1$ and $U_2$ generate a
noncommutative, associative unital $*$-algebra $\alg_\theta$ with
trace called the {\it noncommutative torus}~\cite{Connesbook}. A
typical ``smooth'' element $f\in\alg_\theta$ is of the form
\beq
f=\sum_{\mbf n\in\Z^2}\,f_{\mbf n}~U_1^{n_1}\,U_2^{n_2}
\label{smoothelt}\eeq
where $\{f_{\mbf n}\}\in{\cal S}(\Z^2)$ is a Schwartz sequence. The
trace on $\alg_\theta$ is then defined by
\beq
\ncint f=\Tr(f):=f_{\mbf 0} \ .
\label{ncintdef}\eeq

\subsection{Solution Space \label{SolSpace}}

To determine the structure of the solutions $X_i$ to the quotient conditions
(\ref{quotientconds}), let us momentarily set $\theta=0$. In this case
we can take $\hil={\rm L}^2(\T^2,E)$ to be the Hilbert
space of square-integrable sections of a hermitean vector bundle $E\to\T^2$.
The
operators $U_i$, $i=1,2$ may be represented on $\hil$ as pointwise
multiplication by the functions $\re^{\ri\,\sigma_i}$ where
$\sigma_i\in[0,2\pi)$ are angular coordinates on the
torus. Thus the $U_i$ generate (via Fourier expansion) the commutative
algebra of functions $f$ on a two-torus dual
to $\T^2$. The first equation in (\ref{quotientconds}) is then simply
the Leibnitz rule for a connection on the bundle $E$ given by
$X_i=\ri\,\nabla_i+A_i$, $i=1,2$ and $X_j=A_j$ for $j>2$, where
$\nabla_1,\nabla_2$ specify a constant curvature connection and
$A_i\in\Fun^\infty(\T^2,{\rm End}(E))$. This solution is unique up to
unitary equivalence.

For $\theta\neq0$, the quotient conditions imply that
$X_1,X_2$ are connections on a {\it module} $E$ over the algebra
$\alg_\theta$, while $X_j\in{\rm End}_{\alg_\theta}(E)$ for all $j\neq1,2$. To
describe these solutions explicitly, fix $q\in\Z$ and $p\in\Z/q\,\Z$
with $p,q$ coprime. Set
\beq
E=E_{p,q}:={\rm L}^2(\R)\otimes\C^q
\label{Epqdef}\eeq
and let $\nabla_i$, $i=1,2$ be connections on $E_{p,q}$ of constant
curvature
\beq
\bigl[\nabla_1\,,\,\nabla_2\bigr]=\frac{2\pi\,\ri}{p-q\,\theta}~\one \
.
\label{constcurvcond}\eeq
The separable Hilbert space ${\rm L}^2(\R)$ is the Schr\"odinger
representation of the Heisenberg algebra (\ref{constcurvcond}), which by the
Stone-von~Neumann theorem is the unique irreducible module. The
finite-dimensional Hilbert space $\C^q$ is the irreducible $q\times q$
representation of the Weyl algebra
\beq
\Gamma_1\,\Gamma_2=\re^{2\pi\,\ri\,p/q}~\Gamma_2\,\Gamma_1
\label{Weylalgdef}\eeq
which is uniquely solved (up to unitary equivalence) by ${\rm SU}(q)$
clock and shift matrices $\Gamma_1$ and $\Gamma_2$. Acting on
(\ref{Epqdef}), we then take
\beq
U_i=\exp\bigl(\mbox{$\frac{2\pi\,\ri}q$}\,(p-q\,\theta)\,\nabla_i
\bigr)\otimes\Gamma_i \ , \quad i=1,2 \ .
\label{UiEpqrep}\eeq

This construction stems from the property that projective modules over
the algebra $\alg_\theta$ are classified by the K-theory group ${\rm
  K}_0(\alg_\theta)=\Z\oplus\Z$. If $\theta$ is an
irrational number, then the trace (\ref{ncintdef}) determines an
isomorphism $\Tr:{\rm K}_0(\alg_\theta)\to\Z+\Z\,\theta$ as ordered subgroups
of
$\R$~\cite{lls2,Connesbook}. Stable modules are classified by the
positive cone ${\rm K}_0^+(\alg_\theta)$ defined by positivity of the
Murray-von~Neumann dimension
\beq
\dim(E_{p,q})=\Tr_E(\proj_{p,q})=p-q\,\theta>0
\label{MvNdimEpq}\eeq
where $\proj_{p,q}$ is a hermitean projector
($\proj_{p,q}^2=\proj_{p,q}=\proj_{p,q}^\dag$) such that
$E_{p,q}=\proj_{p,q}\alg_\theta^N$ for some $N$. The
finitely-generated projective module $E_{p,q}$ is called a {\it
  Heisenberg module}.

\subsection{Noncommutative Gauge Theory\label{NCgauge}}

{}From the considerations above, we have thus shown that the general
solutions of the quotient conditions (\ref{quotientconds}) on a
Heisenberg module $E=E_{p,q}$ are given by
\bea
X_i&=&\ri\,\nabla_i+A_i \ , \quad i=1,2 \nonumber\\
X_j&=&A_j \ , \quad j>2
\label{gensolHeisen}\eea
where $A_i\in{\rm End}_{\alg_\theta}(E_{p,q})$ are elements of the
commutant of $\alg_\theta$ in $E_{p,q}$, i.e. the set of operators on
$E_{p,q}$ which commute with the irreducible representation
(\ref{UiEpqrep}). A straightforward computation shows that any such
element admits an expansion
\beq
A_i=\sum_{\mbf n\in\Z^2}A_i({\mbf n})~Z_1^{n_1}\,Z_2^{n_2}
\label{Aigenexp}\eeq
where the endomorphisms $Z_1,Z_2$ are defined by
\bea
(Z_1f_k)(s)&=&\re^{2\pi\,\ri\,s/q}\,f_{k-1}(s) \nonumber\\
(Z_2f_k)(s)&=&\re^{2\pi\,\ri\,k\,a/q}\,f_k\bigl(s+\mbox{$
\frac1{p-q\,\theta}$}\bigr)
\label{Zidef}\eea
for $(f_1,\dots,f_q)\in E_{p,q}={\rm L}^2(\R)\otimes\C^q$. We have
chosen $a,b\in\Z$ to satisfy the first order Diophantine equation
\beq
a\,p+b\,q=1 \ .
\label{dioeq}\eeq

One easily computes that $Z_1,Z_2$ generate another noncommutative
torus $\alg_{\theta'}\cong{\rm End}_{\alg_\theta}(E_{p,q})$ since
\beq
Z_1\,Z_2=\re^{-2\pi\,\ri\,\theta'}~Z_2\,Z_1
\label{ZiNCrel}\eeq
where
\beq
\theta'=\frac{a\,\theta+b}{p-q\,\theta}
\label{thetadual}\eeq
lies in the ${\rm SL}(2,\Z)$ M\"obius orbit of $\theta\in\R$. This
means that $\alg_{\theta'}$ is {\it Morita equivalent} to the algebra
$\alg_\theta$, with $E_{p,q}$ the equivalence bimodule. Finally, setting the
$j>2$ components to~$0$ we find by substituting (\ref{gensolHeisen})
into the matrix model action (\ref{IKKTaction}) the functional
\beq
{ S}(X)={\rm YM}(A):=-\mbox{$\frac1{g^2}$}~\Tr_E\bigl[\ri\,
\nabla_1+A_1\,,\,\ri\,\nabla_2+A_2\bigr]^2 \ .
\label{NCYMaction}\eeq
The noncommutative field theory defined by this action is just
Yang-Mills theory on the Heisenberg module $E=E_{p,q}$.

\section{Matrix Models and Gauge Theory on the Fuzzy
  Torus\label{F-DC}}

In the previous section we began with a matrix model given by a
perfectly well-defined integral over the finite-dimensional space
$\herm(N)\otimes\R^d$. However, for the toroidal compactification of
the model it was necessary to pass to a formal limit whereby the
matrix rank $N\to\infty$ and rewrite objects in terms of (compact)
operators on a separable Hilbert
space. We would now like to understand the origin of this large $N$
limit better and to make it more rigorous. We will
examine how and to what extent the noncommutative torus algebra
$\alg_\theta$ admits representations in terms of finite-dimensional
matrix algebras. This will unveil precisely how a non-perturbative
regularization of noncommutative gauge theory can be obtained and how
the large $N$ continuum limit which removes the regulator $N$ must be
taken.

\subsection{The Eguchi-Kawai Model\label{EK}}

Let us begin by sketching the basic idea behind the finite-dimensional
approximations that we shall construct. If $\theta=\frac MN$ is a
rational number with $M,N\in\N$ coprime, then there exists a
surjective algebra $*$-morphism $\pi:\alg_{M/N}\to\M_N(\C)$ defined on
generators by
\beq
\pi(U_i)=\Gamma_i \ , \quad i=1,2
\label{pimorph}\eeq
where $\Gamma_1$ and $\Gamma_2$ obey the Weyl algebra
$\Gamma_1\,\Gamma_2=\re^{2\pi\,\ri\,M/N}~\Gamma_2\,\Gamma_1$, and thus
generate the finite-dimensional matrix algebra $\M_N(\C)$. In this
context the associative $*$-algebra generated by the Weyl algebra is
sometimes called the {\it fuzzy torus}. Since any
irrational number $\theta$ can be written as the limit of a sequence
of rational numbers, we can anticipate that some sort of limiting
procedure also works at the level of the corresponding algebras. This
issue will be addressed in the next section.

A compact version of the IKKT matrix model
(\ref{IKKTpartfn},\ref{IKKTaction}) can be defined by exponentiating
everything from the Lie algebra to the Lie group. Hence we define the
matrix integral
\beq
{\rm Z}=\int_{{\rm U}(N)\otimes (\Sphere^1)^d}\rd U~\re^{-{\rm S}(U)}
\label{EKpartfn}\eeq
where $\rd U_i$ for each $i=1,\dots,d$ is the left-right invariant
Haar measure on the $N\times N$ unitary group ${\rm U}(N)$ and
\beq
{\rm S}(U)=-\frac1{g^2}~\sum_{1\leq i\neq j\leq d}\,
\Tr\bigl(U_i\,U_j\,U_i^\dag\,U_j^\dag\bigr) \ .
\label{EKaction}\eeq
This unitary matrix model is called the {\it Eguchi-Kawai
  model}~\cite{EK1}. For ``infinitesimal'' values $U_i=\one+\ri\,X_i$,
$X_i\in\herm(N)$ it reduces to the IKKT matrix model. It is the
reduction of ${\rm U}(N)$ Wilson lattice gauge theory on $\Z^d$ to a
single plaquette. The matrix model possesses the gauge symmetry
$U_i\mapsto\Omega\,U_i\,\Omega^{-1}$ with $\Omega\in{\rm U}(N)$.

\subsection{Compact Quotient Conditions\label{F-DQC}}

In the Eguchi-Kawai model there is a perfectly well-defined
finite-dimensional version of the quotient conditions for toroidal
compactification which are obtained via exponentiation of the
constraints (\ref{quotientconds})~\cite{AMNSz1}. They are given by
\bea
\Omega_i\,U_i\,\Omega_i^{-1}&=&\re^{2\pi\,\ri\,r_i/N}~U_i \ ,
\quad r_i\in\Z \ , \quad i=1,2 \nonumber\\
\Omega_i\,U_j\,\Omega_i^{-1}&=&U_j \ , \quad \forall j\neq i \ , \
j=1,\dots,d \ ,  \ i=1,2
\label{compquotient}\eea
where $\Omega_1,\Omega_2\in{\rm U}(N)$. Taking the trace of both sides
of the first equation in (\ref{compquotient}) now only requires that
$\Tr(U_i)=0$ for $i=1,2$.
This truncates the matrix integral (\ref{EKpartfn}) to traceless
unitary matrices, but the equations are still consistent for
finite-dimensional matrices. Similarly to the non-compact case, the
consistency condition generated by the quotient conditions
(\ref{compquotient}) can be chosen to be given by
$\Omega_1\,\Omega_2=\re^{2\pi\,\ri\,l/N}~\Omega_2\,\Omega_1$ for some
$l\in\N$.

We can solve these quotient conditions by introducing discrete
versions of the gauge connections described before acting on
finite-dimensional modules over the matrix algebra $\M_N(\C)$. Let
$N=M\,q$, $M=m\,n\,q$, and represent the noncommutative torus algebra
$\alg_\theta$ in $\M_M(\C)\otimes\M_q(\C)\cong\M_N(\C)$ through
\beq
\Omega_1=\bigl(\Gamma_2\bigr)^m\otimes\bigl(\tilde\Gamma_1^\dag
\bigr)^p \ , \qquad \Omega_2=\bigl(\Gamma_1\bigr)^m
\otimes\tilde\Gamma_2^\dag
\label{Omegaifiniterep}\eeq
where $\Gamma_1\,\Gamma_2=\re^{2\pi\,\ri/M}~\Gamma_2\,\Gamma_1$ and
$\tilde\Gamma_1\,\tilde\Gamma_2=\re^{2\pi\,\ri/q}~\tilde\Gamma_2\,\tilde\Gamma_1$.
Then one has $\Omega_1\,\Omega_2=\re^{2\pi\,\ri\,\theta}~\Omega_2\,\Omega_1$
with
\beq
\theta=\frac pq-\frac m{n\,q} \ .
\label{thetadiscr}\eeq
The commutant of $\alg_\theta$ in $\C^M\otimes\C^q\cong\C^N$ is easily
seen to be generated by the matrices~\cite{AMNSz1}
\beq
Z_1=\bigl(\Gamma_2\bigr)^n\otimes\tilde\Gamma_1^\dag \ , \qquad
Z_2=\bigl(\Gamma_1^\dag\bigr)^n\otimes\bigl(\tilde\Gamma_2\bigr)^a
\label{Zidiscr}\eeq
which obey $Z_1\,Z_2=\re^{2\pi\,\ri\,\theta'}~Z_2\,Z_1$ and thereby
generate another noncommutative torus representation of
$\alg_{\theta'}$ in $\M_N(\C)$ with
\beq
\theta'=\frac n{m\,q}-\frac aq=\frac{a\,\theta+b}{p-q\,\theta} \ .
\label{thetaprimediscr}\eeq

\subsection{Discrete Noncommutative Gauge Theory\label{DiscrNGT}}

We can introduce a fixed discrete gauge ``connection'' on the module
$\C^N$ by setting $r_1=r_2=m\,q$ and defining
\beq
D_1=\Gamma_1^\dag\otimes\one_q \ , \qquad
D_2=\Gamma_2\otimes\one_q \ .
\label{discrconn}\eeq
It has constant curvature given by
\beq
D_1\,D_2=\exp\bigl(\mbox{$\frac{2\pi\,\ri\,q}{p-q\,\theta}$}\,
r_1\,r_2\bigr)~D_2\,D_1 \ .
\label{discrconstcurv}\eeq
Setting $U_j=\one_N$ for $j>2$, the most general solutions to the
compact quotient conditions (\ref{compquotient}) are then given by
\beq
U_i=\tilde U_i\,D_i \ , \quad i=1,2
\label{compquogensol}\eeq
where $\tilde U_i\in\alg_{\theta'}$ are elements of the commutant of
$\alg_\theta$, i.e. $\Omega_j\,\tilde U_i\,\Omega_j^\dag=\tilde U_i$
for $i,j=1,2$. Substituting (\ref{compquogensol}) into the matrix
model action (\ref{EKaction}) thereby leads to a discrete version of
Yang-Mills gauge theory with action
\beq
{\rm S}\bigl(\tilde U\,D\bigr)={\rm W}\bigl(\tilde U\bigr):=
-\mbox{$\frac2{g^2}$}~\Re~\Tr\left[\re^{-2\pi\,\ri/M}~\tilde U_1\,
\bigl(D_1\,\tilde U_2\,D_1^\dag\bigr)\,\bigl(D_2\,\tilde U_1^\dag\,
D_2^\dag\bigr)\,\tilde U_2^\dag\right] \ .
\label{discrYMaction}\eeq
The exact solution of this matrix model
for $n=1$ is given in~\cite{PSz1}.

A basis for the solution space $\alg_{\theta'}$
is provided by the matrices
\beq
J_{\mbf m}=(Z_2)^{m_1}\,(Z_1)^{m_2}~\re^{\pi\,\ri\,\theta'\,m_1\,m_2}
\label{solspacebasis}\eeq
with $\mbf m\in(\Z/m\,q\,\Z)^2$. In terms of this basis we may expand
the discrete gauge fields as
\beq
\tilde U_i=\frac1{(m\,q)^2}~\sum_{\mbf x}\,{\cal U}_i(\mbf x)~
\sum_{\mbf m}\,J_{\mbf m}~\re^{-\frac{2\pi\,\ri}
{m\,q}\,{\mbf m}\wedge\mbf x}
\label{tildeUJexp}\eeq
where $\mbf x=(x_1,x_2)$ with $x_i=0,1,\dots,m\,q-1$. This maps the
discrete gauge theory defined by the action (\ref{discrYMaction}) onto
a noncommutative version of Wilson lattice gauge theory on
$\T^2\cap\Z^2$~\cite{AMNSz1,AMNSz2}.

\section{Large $\mbf N$ Limit\label{AF-ARA}}

We will now describe precisely the large $N$ limit of the
finite-dimensional approximation of Section~3 which leads
back to the original noncommutative gauge theory constructed in
Section~2. We will focus on how to do this at a purely algebraic
level, and then discuss some of the important topological consequences
of the construction.

\subsection{AF-Algebras\label{AF-Alg}}

The idea behind the construction of this section is to take the
``large $N$ limit'' by embedding the noncommutative torus algebra
$\alg_\theta$ into an {\it approximately finite-dimensional (AF)}
algebra
\beq
A_\infty=\lim_{\stackrel{\scriptstyle\longrightarrow}{n\in\N_0}}\,A_n
\label{Ainftydef}\eeq
defined as the norm closure of the
inductive limit of an inductive system
\beq
A_0~\stackrel{\rho_1}{\longrightarrow}~A_1~\stackrel{\rho_2}
{\longrightarrow}~A_2~\stackrel{\rho_3}{\longrightarrow}~\cdots~
A_n~\stackrel{\rho_{n+1}}{\longrightarrow}~\cdots
\label{indsystem}\eeq
where each $A_n$ is a finite-dimensional $C^*$-algebra with the usual
operator norm $\|-\|_{A_n}$, and $\rho_n$ are
injective $*$-morphisms. The inductive limit (\ref{Ainftydef}) is a
$C^*$-algebra with norm given by
\beq
\bigl\|(f_n)_{n\in\N_0}\bigr\|_{A_\infty}:=\lim_{n\to\infty}\,\bigl\|f_n
\bigr\|_{A_n}
\label{Ainftynorm}\eeq
with $f_n\in A_n$. We can use the embeddings (\ref{indsystem}) to
identify $A_n$ with a subalgebra of $A_{n+1}$ as
\beq
A_n=\bigoplus_{j=1}^{l_n}\,\M_{d_j^{(n)}}(\C)\cong
\bigoplus_{k=1}^{l_{n+1}}~\bigoplus_{j=1}^{l_n}\,\M_{d_j^{(n)}}
(\C)\otimes\C^{N_{kj}}
\label{Anembed}\eeq
with
$\bigoplus_{j=1}^{l_n}\,\M_{d_j^{(n)}}(\C)\otimes
\C^{N_{kj}}\subset\M_{d_k^{(n+1)}}(\C)$.
The non-negative integers $N_{kj}$ must obey the consistency
conditions
\beq
\sum_{j=1}^{l_n}\,N_{kj}\,d_j^{(n)}=d_k^{(n+1)}
\label{conscondembed}\eeq
which define a collection of partial embeddings
$d_j^{(n)}\searrow^{\!N_{kj}}d_k^{(n+1)}$.

The important point is that the algebra $\alg_\theta$ is {\it not} an
AF-algebra. This can be seen, for instance, at the level of
K-theory. The degree~$1$ K-theory group of any finite-dimensional
algebra is always trivial, and since K-theory is covariant under
inductive limits one has $\K_1(A_\infty)=0$. On the other hand, the
noncommutative torus has non-trivial K-theory group
$\K_1(\alg_\theta)=\Z\oplus\Z$, with generators the unitary
equivalence classes of $U_1$ and $U_2$. Thus, at the
level of zero-dimensional matrix models, the large $N$ limit can at
best be defined by an embedding $\alg_\theta\hookrightarrow
A_\infty$. This has the consequence of making the large $N$ limit rather
complex, a property also witnessed of the complicated double-scaling
continuum limit of the noncommutative lattice gauge theory of the
previous section~\cite{AMNSz1,AMNSz2}. We will see a cleaner way to
treat the large $N$ limit later on via one-dimensional matrix
models. An approximation of the noncommutative torus by fuzzy tori with
respect to the quantum Gromov-Hausdorff metric has been recently
constructed in~\cite{Latre1}.

\subsection{Rational Approximation\label{RatApprox}}

To realize this construction explicitly~\cite{LLSz1,PV1}, we expand
the irrational number $\theta\in\R\setminus\Q$ into simple continued
fractions as~\cite{HardyWright1}
\beq
\theta=\lim_{n\to\infty}\,\theta_n
\label{thetalim}\eeq
with the $n$-th convergent of the expansion given by
\beq
\theta_n=\frac{p_n}{q_n}:=c_0+\frac1{\ds c_1+\frac1{\ds c_2+\frac1{
\ds\ddots~ c_{n-1}+\frac1{c_n}}}}
\label{contfracexp}\eeq
where $c_k\in\N$ for $k\geq1$, $c_0\in\Z$, and the coprime integers
$p_n,q_n$ can be computed from the recursion relations
\bea
p_0&=&c_0 \ , \quad p_1~=~c_0\,c_1+1 \ , \quad p_n~=~c_n\,
p_{n-1}+p_{n-2} \nonumber\\ q_0&=&1 \ , \ \ \ \ \ \
q_1~=~c_1 \ \ \ \qquad  \ , \quad
q_n~=~c_n\,q_{n-1}+q_{n-2}
\label{pnqnrecurs}\eea
for $n\geq2$. It follows from these relations that the positive
sequences $\{q_n\}$ and $\{|p_n|\}$ are increasing with
$q_n,|p_n|\to\infty$ as $n\to\infty$. The desired finite-dimensional
algebra $A_n$ at level $n\in\N_0$ is then given by
\beq
A_n:=\M_{q_n}(\C)\oplus\M_{q_{n-1}}(\C)
\label{Anexpldef}\eeq
with the embeddings $\rho_n:A_{n-1}\hookrightarrow A_n$
defined by
\beq
\mbf M\oplus\mbf N~\stackrel{\rho_n}{\longmapsto}~\bigl(\mbf M^{\oplus c_n}
\oplus\mbf N\bigr)\oplus\mbf M
\label{embexpldef}\eeq
for $\mbf M\in\M_{q_{n-1}}(\C)$ and $\mbf N\in\M_{q_{n-2}}(\C)$. At each
finite level $n$, let $U_1^{(n)},U_2^{(n)}$ be the generators of the
noncommutative torus algebra $\alg_{\theta_n}$ obeying
\beq
U_1^{(n)}\,U_2^{(n)}=\re^{2\pi\,\ri\,\theta_n}~U_2^{(n)}\,
U_1^{(n)} \ .
\label{levelngens}\eeq

\subsection{Matrix Approximation\label{MatApprox}}

We can finally derive the matrix approximation to the algebra
$\alg_\theta$ which rigorously accomplishes the desired large $N$
limit of the unitary matrix model. As mentioned in the previous
section, for each $n$ there is a surjective algebra homomorphism
$\pi_n:\alg_{\theta_n}\to\M_{q_n}(\C)$ given by 
\beq
\pi_n\bigl(U_i^{(n)}\bigr)=\Gamma_i^{(n)} \ , \quad i=1,2
\label{pinhomo}\eeq
with $\Gamma_1^{(n)},\Gamma_2^{(n)}$ the $q_n\times q_n$ clock and
cyclic shift matrix generators of $\M_{q_n}(\C)$ which obey the Weyl
algebra
\beq
\Gamma_1^{(n)}\,\Gamma_2^{(n)}=\re^{2\pi\,\ri\,p_n/q_n}~\Gamma_2^{(n)}\,
\Gamma_1^{(n)} \ .
\label{GammanWeyl}\eeq
Then the subalgebra
$\pi_n(\alg_{\theta_n})\oplus\pi_{n-1}(\alg_{\theta_{n-1}})\subset
A_n$ is a finite-dimensional approximation of $\alg_\theta$ in the
following sense~\cite{LLSz1}. Since~\cite{PV1}
\beq
\lim_{n\to\infty}\,\bigl\|\rho_n\bigl(\Gamma_i^{(n-1)}\oplus
\Gamma_i^{(n-2)}\bigr)-\Gamma_i^{(n)}\oplus\Gamma_i^{(n-1)}
\bigr\|_{A_n}=0
\label{rhoGammanlim}\eeq
for $i=1,2$, it follows that there exist unitary operators $U_i\in
A_\infty$ which are limits of sequences of finite-rank operators in
the inductive limit (\ref{Ainftydef}) with respect to the induced operator
norm (\ref{Ainftynorm}) on $A_\infty$, and which obey the defining
relation (\ref{NCtorusdef}) of the noncommutative torus. Thus there
exists a unital injective $*$-morphism $\rho:\alg_\theta\to
A_\infty$.

It is in this sense that the elements of the algebra $\alg_\theta$ may
be ``approximated'' by sufficiently large finite-dimensional matrices
and the large $N$ limit thus taken, since for $n$ sufficiently large
the generators $\Gamma_i^{(n)}$ are well approximated by the images
under the injection $\rho_n$ of the matrices $\Gamma_i^{(n-1)}$
generating $\alg_{\theta_{n-1}}$. The embeddings $\rho_n:A_{n-1}\to
A_n$ above are completely characterized by the sequence of partial
embeddings $\{c_n\}_{n\in\N_0}$ associated with the positive maps
$\varphi_n:\Z^2\to\Z^2$ given by
\beq
\left({}^{~\,q_n}_{q_{n-1}}\right)=\varphi_n\left({}^{q_{n-1}}_{q_{n-2}}
\right) \ , \qquad \varphi_n=\biggl({}^{c_n}_{~1}~{}^1_0\biggr) \ .
\label{posmaps}\eeq
It follows that the K-theory group $\K_0(A_\infty)$ can be obtained as
the inductive limit of the inductive system of ordered groups
$\{\varphi_n:\K_0(A_{n-1})\to\K_0(A_n)\}_{n\in\N_0}$. Since
$\K_0(A_n)=\Z\oplus\Z$ (with the canonical ordering
$\K_0^+(A_n)=\N\oplus\N$) for all $n\geq0$, there is
an isomorphism of ordered groups $\K_0(A_\infty)\cong\Z+\Z\,\theta$
with positive cone
$\K_0^+(A_\infty)=\{(p,q)\in\Z^2~;~p-q\,\theta>0\}$. This coincides
with the K-theory of the noncommutative torus algebra $\alg_\theta$.

The sets $\Z+\Z\,\theta$ and $\Z+\Z\,\theta'$ are isomorphic as ordered groups
if and only if the irrational numbers $\theta,\theta'$ lie in the same
${\rm SL}(2,\Z)$ orbit as in (\ref{thetadual})~\cite{HardyWright1},
i.e. the algebras $\alg_\theta$ and $\alg_{\theta'}$ are Morita
equivalent. Equivalently, the continued fraction expansions of
$\theta$ and $\theta'$ have the same ``tails''. This has two important
consequences~\cite{LLSz1}. Firstly, Morita equivalent tori have the
same K-theory group. Secondly, Morita equivalent noncommutative tori can be
embedded in the {\it same} AF-algebra $A_\infty$ (up to isomorphism),
because their sequences of embeddings are the same up to a finite
number of terms.

\section{Noncommutative Solitons\label{NCSolitons}}

Given the complexity of the large $N$ limit required of the
zero-dimensional matrix models of noncommutative gauge theory, it is
desirable both physically and mathematically to seek alternative
matrix regularizations for which a simpler continuum limit exists. We
will now describe precisely how to do this using one-dimensional
matrix models, i.e. matrix quantum mechanics. The large $N$ limit of these
matrix models does not require a complicated double-scaling, nor is it
the conventional 't~Hooft planar limit. The matrix approximation is
intimately related to the regularization of generic noncommutative
field theories by means of solitons on the noncommutative torus, to
which the present section is devoted. This formalism also has the
virtue of making contact with the relationship between
noncommutative field theory and the dynamics of D-branes in string
theory.

\subsection{D-Branes and Solitons\label{NCSolPlane}}

Let us begin by briefly describing the somewhat simpler situation of
solitons on the noncommutative plane, viewed in terms of operators on
the Fock module (Schr\"odinger representation) over the Heisenberg
algebra. Open string field theory on this module is described by a
potential energy functional $V$. Restrict to static solutions of the
equations of motion. Suppose that $V=V(\tach^2)$ is an even functional
of hermitean elements $\tach$ corresponding to tachyon fields. Then
the equations of motion are
$\tach\,V'(\tach^2)=0$, which for polynomial functions $V$ can be
solved in terms of projections $\tach=\tach^2$. Let
$\{|n\rangle\}_{n\in\N_0}$ be the standard orthonormal number basis
for the Fock space. Then the basic rank $N$ projector $\tach_N$,
having $\Tr(\tach_N)=N$, is given by
\beq
\tach_N=\sum_{n=0}^{N-1}\,|n\rangle\langle n|
\label{tachNdef}\eeq
and it describes $N$ D0-branes sitting inside a D2-brane. In
deformation quantization~\cite{Szrev1}, the Wigner function
corresponding to the operator $\tach_1$ is a gaussian field on $\R^2$
centred about the origin with width proportional to $\theta^{-1/2}$,
and hence corresponds to a solitonic ``lump''. This is the basic GMS
soliton~\cite{GMS1}. More generally, if
$V=V(\tach\,\tach^\dag-\one)+V(\tach^\dag\,\tach-\one)$ is a
functional of generically complex operators $\tach$, then the static
equations of motion are solved by partial isometries
$\tach=\tach\,\tach^\dag\,\tach$. Equivalently, the operators
$\tach^\dag\,\tach$ and $\tach\,\tach^\dag$ are projections. Such
tachyon fields describe brane-antibrane systems and the basic partial
isometry $\tach={\sf S}$ of the Fock module is provided by the
standard shift operator
\beq
{\sf S}=\sum_{n\in\N_0}\,|n+1\rangle\langle n| \ .
\label{shiftopdef}\eeq

The analogous quantities on the noncommutative torus are much
richer and intricate, and it is the purpose of the remainder of this
section to demonstrate how they are constructed. In the next
section we will then show that fields on the noncommutative torus
$\alg_\theta$ are expandable in a basis of projection and partial
isometry solitons. The solitons generate {\it subalgebras}
$\alg_n\subset\alg_\theta$ which are isomorphic to two copies of the
algebra of matrix-valued functions on a circle of the form
$\alg_n\cong\M_{q_{2n}}(\Sphere^1)\oplus\M_{q_{2n-1}}(\Sphere^1)$,
and for which the convergence to $\alg_\theta$ as $n\to\infty$ is
``exact'' in the sense that $\alg_\theta$ is the inductive
limit~\cite{lls2,EE1}
\beq
\alg_\theta=\lim_{\stackrel{\scriptstyle\longrightarrow}{n\in\N_0}}\,
\alg_n \ .
\label{algthetaindlim}\eeq
It follows that any field theory on $\alg_\theta$ is a {\it matrix
  quantum mechanics} with a much simpler large $N$ limit than
before.

\subsection{Powers-Rieffel Projections\label{P-RProjs}}

Given the generators $U_1,U_2$ of $\alg_\theta$ obeying
(\ref{NCtorusdef}) and the continued fraction expansion
(\ref{thetalim},\ref{contfracexp}), we define two {\it towers} of
projections
\bea
\proj_n&=&U_1^{-q_{2n-1}}\,\hat g_n+\hat f_n+\hat g_n\,U_1^{q_{2n-1}}
\nonumber\\ \proj_n'&=&U_2^{q_{2n}}\,\hat g_n'+\hat f_n'+\hat g_n'\,
U_2^{-q_{2n}} \ .
\label{2seqprojs}\eea
To ease notation, we will drop the sequence labels $n\in\N_0$ until we
look at the $n\to\infty$ limits explicitly, and denote $q:=q_{2n}$ and
$q':=q_{2n-1}$. Then $q,q'\to\infty$ in the limit we are
interested in. The algebra element $\hat f=\rho(f)$ is in correspondence with
a function $f\in\Fun^\infty(\Sphere^1)$ through the map
$\rho:\Fun^\infty(\Sphere^1)\to\alg_\theta$ defined on generators by
$\rho(z):=U_1$, where $z$ is the coordinate of the circle
$\Sphere^1$. Similarly, $\hat f'=\rho'(f'\,)$ corresponds to a
function $f'$ on $\Sphere^1$ through the dual map
$\rho':\Fun^\infty(\Sphere^1)\to\alg_\theta$ given by $\rho'(z):=U_2$,
and analogous statements are true of $\hat g,\hat g'$. The
traces of these elements are given by $\ncint\hat f=f(1)$, and so on.

The functions $f,g,f',g'\in\Fun^\infty(\Sphere^1)$ take values in
the interval $[0,1]$ and are called {\it bump functions} because they
are zero almost everywhere on $\Sphere^1$~\cite{lls2}. They are chosen so
that the elements (\ref{2seqprojs}) satisfy three basic requirements:
(a) They are projectors, $\proj^2=\proj$ and
$\proj^{\prime\,2}=\proj'$; (b) They have ranks
$\ncint\proj=p'-q'\,\theta=:\beta$ and
$\ncint\proj'=-(p-q\,\theta)=:\beta'$; and (c) They have Chern numbers
$c_1(\proj)=-q'$ and $c_1(\proj'\,)=q$. This fixes the K-theory classes
of the projectors (\ref{2seqprojs}) which are interpreted as
$({\rm D}2,{\rm D}0)$-brane charges $(p',-q')$ and $(-p,q)$
respectively in open string field
theory~\cite{Bars1}. The non-trivial generator of
$\K_0(\alg_\theta)$ has charge $(0,1)$ and is called the {\it
  Powers-Rieffel projector}~\cite{Rieffel1}.

\subsection{Orthogonal Projections\label{OrthoProjs}}

Let us focus for the moment on the first tower of projectors
$\proj$ in (\ref{2seqprojs}). In deformation
quantization~\cite{Szrev1}, the Wigner function corresponding to
$\proj$ is not a lump as in the case of the noncommutative plane, but
rather exihibits stripe patterns on the torus $\T^2$ with area proportional to
$\beta$~\cite{lls2}. One set of stripes displays periodic lumps with
period $q'$, which is a manifestation of the UV/IR mixing phenomenon
in noncommutative field theory since the size of the soliton grows
with its oscillation period. We can now ``translate'' the projector
$\proj$ along the first cycle of $\T^2$ through the outer automorphism
$\alpha:\alg_\theta\to\alg_\theta$ defined by
\beq
\alpha(U_2)=\re^{2\pi\,\ri\,p/q}~U_2 \ , \quad \alpha(U_1)=U_1 \ .
\label{alphadef}\eeq
Iterating, we may then define a new set of projections for
$i=1,\dots,q$ by
\beq
\proj^{ii}:=\alpha^{i-1}(\proj) \ .
\label{projiidef}\eeq
They form a system of mutually orthogonal projection operators with
\beq
\proj^{ii}\,\proj^{jj}=\delta_{ij}~\proj^{ii} \ .
\label{orthoprojsys}\eeq

It is convenient in this construction to represent the algebra on the
GNS representation space $\hil:={\rm L}^2(\alg_\theta\,,\,\ncint)$. The
images of the projectors (\ref{projiidef}) then define the {\it
  Chan-Paton subspaces} $\hil_i:=\im(\proj^{ii})\subset\hil$ with
${\rm End}(\hil_i)=\proj^{ii}\,\alg_\theta\,\proj^{ii}$ the algebra of
open string modes ending on the D-brane state determined by
$\proj^{ii}$. One has $\proj^{ii}|_{\hil_i}=\one$ and
$\hil_i\subset\ker(\proj^{jj})$ for~$j\neq i$.

\subsection{Partial Isometries\label{PartIso}}

Let us define the operator
\beq
\Pi^{21}:=\proj^{22}\,U_1\,\proj^{11} \ .
\label{Pi21def}\eeq
It can be regarded as a closed bounded operator
$\Pi^{21}:\hil_1\to\hil_2$, but it is not an isometry since
$(\Pi^{21})^\dag\,\Pi^{21}\neq\one$. On $\hil$ it admits a polar
decomposition given by~\cite{Calg1}
\beq
\Pi^{21}:=\proj^{21}\,\bigl|\Pi^{21}\bigr|
\label{polardecomp}\eeq
where $|\Pi^{21}|$ is a hermitean operator and
$\proj^{21}\in\alg_\theta$ is a partial isometry with~\cite{EE1}
\beq
\lim_{n\to\infty}\,\bigl\|\Pi_n^{21}-\proj_n^{21}\bigr\|_\hil=0 \ .
\label{partisolim}\eeq

For $i=1,\dots,q-2$ the translated partial isometries
\beq
\proj^{i+2,i+1}:=\alpha^i\bigl(\proj^{21}\bigr) \ , \quad
\proj^{ji}:=\bigl(\proj^{ij}\bigr)^\dag
\label{partisotransl}\eeq
satisfy the {\it matrix unit relations}
\beq
\proj^{ij}\,\proj^{kl}=\delta_{jk}~\proj^{il} \ .
\label{matrixunits}\eeq
These relations can be used to generate a set of $q^2$ operators
$\proj^{ij}:\hil_j\to\hil_i$,
i.e. $\proj^{ij}\in\proj^{ii}\,\alg_\theta\,\proj^{jj}$. In a similar
fashion, from the second tower we can construct $q^{\prime\,2}$ matrix
units $\proj^{\prime\,i'j'}$ which are orthogonal to $\proj^{ij}$.

\section{Noncommutative Field Theory as Matrix Quantum
  Mechanics\label{NCFTMQM}}

We will now use the systems of projections and partial isometries
above to construct the desired one-dimensional matrix
model~\cite{lls2}. We will first build the subalgebras $\alg_n$ and
state the basic matrix approximation theorem at the algebraic
level. Then we demonstrate how to transcribe generic field theory
actions on $\alg_\theta$ into matrix quantum mechanics which can serve
as precise non-perturbative regularizations of the continuum field
theories with tractable large $N$ limits.

\subsection{The Matrix Approximation\label{MatrixSubalg}}

The collection of operators $\{\proj^{ij}\}$ do not quite close a
$q\times q$ matrix algebra, because
\beq
\proj^{1q}:=\proj^{12}\,\proj^{23}\cdots\proj^{q-1,q}\neq\alpha^{q-1}
\bigl(\proj^{21}\bigr) \ .
\label{proj1q}\eeq
However, both operators in (\ref{proj1q}) are isometries on
$\hil_q\to\hil_1$ and consequently are related as
\beq
\alpha^{q-1}\bigl(\proj^{21}\bigr)=z~\proj^{1q}
\label{proj211qrel}\eeq
where $z$ is a unitary operator on $\hil_1$, i.e. a partial isometry
on the whole of the Hilbert space $\hil$. We may regard $z$ as the
generator of the algebra $\Fun^\infty(\Sphere^1)$. Then the operators
$\{\proj^{ij},z\}$ close a subalgebra of $\alg_\theta$ which is
naturally isomorphic to the algebra $\M_q(\Sphere^1)$ of $q\times q$
matrix-valued functions on a circle. An analogous construction in the
second tower gives a collection $\{\proj^{\prime\,i'j'},z'\,\}$, and
combining the two towers thus gives the matrix subalgebras
\beq
\alg_n\cong\M_q\bigl(\Sphere^1\bigr)\oplus\M_{q'}\bigl(\Sphere^1
\bigr)\subset\alg_\theta \ .
\label{matrixsubalgdef}\eeq
The important point here is that $\alg_n$ is a {\it subalgebra} of
$\alg_\theta$.

By using the continued fraction expansion
(\ref{thetalim},\ref{contfracexp}) it is possible to define a system
of embeddings on the sequence of subalgebras
$\{\alg_n\}_{n\in\N_0}$ and realize the noncommutative torus algebra
$\alg_\theta$ as an inductive limit (\ref{algthetaindlim})~\cite{lls2}. To
describe the convergence theorem explicitly, we define the operators
\beq
\mbf U_1=\Gamma_1^{(q)}\oplus\Gamma_2^{(q'\,)}(z'\,) \ , \qquad
\mbf U_2=\Gamma_2^{(q)}(z)\oplus\Gamma_1^{(q'\,)} \ .
\label{mbfUdef}\eeq
The operator $\Gamma_1^{(q)}$ is, with respect to the system of matrix
units $\proj^{ij}\in\alg_\theta$, the standard $q\times q$ clock matrix
in the $q$-th root of unity $\re^{2\pi\,\ri\,\theta_{2n}}$. The operator
$\Gamma_2^{(q)}(z)$ is the same as the standard shift matrix except
that its component multiplying the matrix unit $\proj^{q1}$ is $z$. It
coincides with the standard cyclic shift matrix
$\Gamma_2^{(q)}=\Gamma_2^{(q)}(1)$ at $z=1$. One still has the usual
Weyl algebra (\ref{GammanWeyl}), and also
$\bigl(\Gamma_1^{(q)}\bigr)^q=\one_q,\bigl(\Gamma_2^{(q)}(z)
\bigr)^q=z~\one_q$. Completely
analogous relations hold for the second tower, and it follows that the
elements $\mbf U_1,\mbf U_2$ generate the matrix subalgebra
(\ref{matrixsubalgdef}). We may define a restriction map
$\gamma_n:\alg_\theta\to\alg_n$ on generators by
\beq
\gamma_n(U_i)=\mbf U_i \ , \quad i=1,2 \ .
\label{gammandef}\eeq
Then for any element $f\in\alg_\theta$ the image
$\gamma_n(f):=\bff\oplus\bff'\in\alg_n$ converges to $f$ in the sense
that~\cite{lls2}
\beq
\lim_{n\to\infty}\,\bigl\|f-\gamma_n(f)\bigr\|_\hil=0 \ .
\label{gammanconv}\eeq
The mapping $\gamma_n$, which is {\it not} an algebra homomorphism,
defines the {\it soliton expansion} of noncommutative fields.

\subsection{The One-Dimensional Matrix Model\label{1DMM}}

We will now describe how to construct matrix model actions which
approximate generic field theories on the noncommutative torus~\cite{lls2}.
Let
$\partial_i:\alg_\theta\to\alg_\theta$ be the outer derivations
representing the action of the translation group of $\T^2$ which are
defined on generators by
\beq
\partial_i(U_j)=2\pi\,\ri\,\delta_{ij}\,U_j \ , \quad i,j=1,2 \ .
\label{derivdef}\eeq
Let $\Phi$ be a collection of fields in $\alg_\theta$ with lagrangian
$\Lcal(\Phi,\partial_i(\Phi))\in\alg_\theta$. Field theory on
$\alg_\theta$ is then defined by the action
\beq
S(\Phi)=\ncint\Lcal\bigl(\Phi\,,\,\partial_i(\Phi)\bigr) \ .
\label{SPhidef}\eeq
In the setting of Sections~3 and~4, we use the map
$\pi_n:\alg_\theta\to\M_{q_n}(\C)$ defined by
$\pi_n(U_i)=\Gamma_i^{(q_n)}$ (c.f.~eq.~(\ref{pinhomo})) and the
discrete derivatives (\ref{discrconn}) to approximate this field
theory by a zero-dimensional matrix model with action~\cite{AMNSz2}
\beq
S_n^{(0)}(\Phi)=\Tr\,\Lcal\bigl(\pi_n(\Phi)\,,\,D_i\,\pi_n(\Phi)\,
D_i^\dag-\pi_n(\Phi)\bigr) \ .
\label{0dimapprox}\eeq
We will now find a matrix quantum mechanics action $S_n^{(1)}(\Phi)$
corresponding to the approximation $\gamma_n:\alg_\theta\to\alg_n$
defined by (\ref{gammandef}).

Let us begin by describing how to transcribe the trace in
(\ref{SPhidef}). On $\alg_\theta$ it is defined on generic elements
(\ref{smoothelt}) by $\ncint
U_1^{n_1}\,U_2^{n_2}=\delta_{n_1,0}\,\delta_{n_2,0}$. On the matrix
subalgebra $\alg_n$, it is possible to work out
$\ncint\bfu_1^{n_1}\,\bfu_2^{n_2}$ using the trace properties
$\ncint\proj^{ij}=\beta~\delta_{ij}$ of the matrix units. Using the rapid
decay property of Schwartz sequences, in the limit $n\to\infty$ one
finds that a good approximation is given by the expected definition
\beq
\ncint\gamma_n(f)=\beta\,\int_0^1\rd\tau~\Tr\,\bff(\tau)+
\beta'\,\int_0^1\rd\tau'~\Tr\,\bff'(\tau'\,)
\label{ncintapprox}\eeq
where we have parametrized the circles in the two towers by
$z=\re^{2\pi\,\ri\,\tau}$ and $z'=\re^{2\pi\,\ri\,\tau'}$ with
$\tau,\tau'\in[0,1)$.

The approximation of the derivations (\ref{derivdef}) is much more
involved. Let us focus on the first tower in (\ref{matrixsubalgdef})
and expand the component of $\gamma_n(f)$ in this tower as
\beq
\bff=\sum_{l,m=1}^q~\sum_{k\in\Z}\,\phi_{lm;k}~z^k~\Gamma_1^l\,
\Gamma_2(z)^m
\label{bffexp}\eeq
with $\phi_{lm;k}:=\sum_{r\in\Z}\,f_{(l+r\,q,m+k\,q)}$. By considering
the projection $\gamma_n(\partial_i(f))$ on $\alg_\theta\to\alg_n$ and
using the Schwartz property in the limit $q\to\infty$, we define the
operators
\bea
\Delta_1\bff(z)&:=&2\pi\,\ri\,\sum_{l,m=1}^q~\sum_{k\in\Z}\,l\,
\phi_{lm;k}~z^k~\Gamma_1^l\,\Gamma_2(z)^m \nonumber\\
\Delta_2\bff(z)&:=&2\pi\,\ri\,\sum_{l,m=1}^q~\sum_{k\in\Z}\,
(m+k\,q)\,\phi_{lm;k}~z^k~\Gamma_1^l\,\Gamma_2(z)^m \ .
\label{nablaapproxdef}\eea
These operators converge to the derivations $\partial_i$ in the limit
$n\to\infty$. At finite $n$ they satisfy an ``approximate'' Leibnitz
rule, in the sense that they only become derivations at
$n\to\infty$. To write these as operators acting on the expansion of
$\bff$ in the system of matrix units $\proj^{lm}$, one now needs to
compute the change of orthogonal bases between $\proj^{lm}$ and
$\Gamma_1^l\,\Gamma_2(z)^m$~\cite{lls2}.

By performing an identical analysis in the second tower, in this way
one finds that the action functional (\ref{SPhidef}) is well
approximated by the one-dimensional matrix model with action
\bea
S_n^{(1)}(\Phi)&=&\beta\,\int_0^1\rd\tau~\Tr\,\Lcal\bigl(\bfphi(\tau)\,,\,
\Delta_i\bfphi(\tau)\bigr)\nonumber\\
&&+\,\beta'\,\int_0^1\rd\tau'~\Tr\,\Lcal\bigl(
\bfphi'(\tau'\,)\,,\,\Delta_i\bfphi'(\tau'\,)\bigr)
\label{Sn1Phidef}\eea
where $\gamma_n(\Phi):=\bfphi(\tau)\oplus\bfphi'(\tau'\,)$ and
\bea
\Delta_1\bfphi(\tau)&=&\Sigma\bfphi(\tau) \ , \qquad
\Delta_1\bfphi'(\tau'\,)~=~q'\,\dot\bfphi{}'(\tau'\,)+\bigl[
\Xi'\,,\,\bfphi'(\tau'\,)\bigr] \nonumber\\
\Delta_2\bfphi(\tau)&=& q\,\dot\bfphi(\tau)+\bigl[
\Xi\,,\,\bfphi(\tau)\bigr]\ , \qquad
\Delta_2\bfphi'(\tau'\,)~=~\Sigma'\bfphi'(\tau'\,)
\label{nablafinal}\eea
with the dots denoting a derivative with respect to $\tau$ or
$\tau'$. The $q\times q$ matrix
\beq
\Xi_{lm}=2\pi\,\ri\,m~\delta_{lm} \ , \quad 1\leq l,m\leq q
\label{infclock}\eeq
is an ``infinitesimal'' version of the clock matrix
$\bigl(\Gamma_1^{(q)}\bigr)^\dag$. The antihermitean operator $\Sigma$
is defined by
$(\Sigma\bff)(\tau)_{lm}=\sum_{s,t}\,\Sigma(\tau)_{lm,st}\,\bff(\tau)_{st}$
with
\beq
\Sigma(\tau)_{lm,st}=-\frac{2\pi\,\ri}q\,\sum_{s'=1}^q\,s'~
\re^{2\pi\,\ri\,s'\,(l-s)\,\theta_{2n}}~
\left\{\begin{matrix}\scriptstyle
\delta_{t,s+l-m} \ , \ \scriptstyle m\leq l \ , \ 
\scriptstyle 1\leq s\leq q+m-l\\
\scriptstyle\delta_{t,s+l-m-q}~\re^{-2\pi\,\ri\,\tau} \ , \
\scriptstyle j<i \ , \ \scriptstyle
q+m-l+1\leq s\leq q\end{matrix} \right.
\eeq
and it can be regarded as an ``infinitesimal'' version of the shift
matrix $\Gamma_2^{(q)}$.

The forms (\ref{nablafinal}) (along with (\ref{mbfUdef})) of the
derivation $\Delta$ illustrate the
role of the two towers in this matrix approximation. The roles of the
components of $\Delta$ are interchanged between the two towers, and
acting on Schwartz sequences they ``compensate'' each other in the
$n\to\infty$ limit. Because of the inductive limit property
(\ref{algthetaindlim}), the convergence of (\ref{Sn1Phidef}) to the
original continuum action (\ref{SPhidef}) is ``exact'' and the large
$n$ limit may be taken directly. This is in marked
contrast to the large $n$ limit required of (\ref{0dimapprox}) which
involves a complex double scaling limit via an embedding of the
noncommutative torus into a homotopically equivalent AF-algebra
$A_\infty$. The zero-dimensional matrix model has a clear geometric
origin through toroidal compactification, while the one-dimensional
matrix model has a nice physical interpretation. The mapping
$f\mapsto\gamma_n(f)$ truncates the
infinite number of image D-branes living on the covering space of
$\T^2=\R^2/\Z^2$ to a finite number $q\,q'$, corresponding to the
physical open string modes which are invariant under the action of the
truncated momentum lattice $(\Z/q\,q'\,\Z)^2$.

\section*{Acknowledgements}

The author would like to thank the organisors and participants of the
workshop for having provided a pleasant and stimulating atmosphere. It is a
pleasure to thank G.~Goldin, V.~Mathai, H.~Steinacker and T.~Voronov
for enjoyable discussions. This work was supported in part by PPARC
Grant PPA/G/S/2002/00478 and by the EU-RTN Network Grant
MRTN-CT-2004-005104.

\aut{Department of Mathematics\\ and\\
Maxwell Institute for Mathematical Sciences\\
Heriot-Watt University\\
Colin Maclaurin Building\\ Riccarton, Edinburgh EH14 4AS\\
United Kingdom\\
{\it E-mail address}:\\
 {\tt R.J.Szabo@ma.hw.ac.uk}}
{}

\label{last}
\end{document}